\documentclass{aa}

\usepackage{natbib}

\begin{document}

\title{The MultiSite Spectroscopic Telescope campaign: 2\,m spectroscopy of
  the V361\,Hya variable PG\,1605+072}

\author{S.~J. O'Toole
  \inst{1,2}
  \and U. Heber
  \inst{1}
  \and C.~S. Jeffery
  \inst{3}
  \and S. Dreizler
  \inst{4,5}
  \and S.~L. Schuh
  \inst{4,5}
  \and V.~M. Woolf
  \inst{3,6}
  \and S. Falter
  \inst{1,8}
  \and E.~M. Green
  \inst{7}
  \and B.-Q. For
  \inst{7}
  \and E.~A. Hyde
  \inst{7}
  \and H. Kjeldsen
  \inst{9}
  \and T. Mauch
  \inst{2}
  \and B.~A. White
  \inst{7}}

\institute{Dr Remeis-Sternwarte, Astronomisches Institut der
  Universit\"at Erlangen-N\"urnberg, Sternwartstr. 7, Bamberg D-96049,
  Germany
  \and School of Physics, University of Sydney, NSW 2006, Australia
  \and Armagh Observatory, College Hill, Armagh BT61 9DG, UK
  \and Institut f\"ur Astrophysik, Universit\"at G\"ottingen,
              Friedrich-Hund-Platz 1, D-37077 G\"ottingen, Germany
  \and Institut f\"ur Astronomie und Astrophysik, Universit\"at
  T\"ubingen, Sand 1, T\"ubingen D-72076, Germany
  \and Department of Physics, University of Nebraska at Omaha, 6001
  Dodge St, Omaha NE 68182, USA
  \and Steward Observatory, University of Arizona, 933 North Cherry Avenue,
  Tucson, AZ 85721, USA
  \and Max Planck Institut f\"ur Astronomie, K\"onigstuhl 17,
  Heidelberg D-69117, Germany
  \and Teoretisk Astrofysik Center, Danmarks Grundforskningsfond, 8000 Aarhus
  C, Denmark
}
 
\offprints{Simon O'Toole,
\email{otoole@sternwarte.uni-erlangen.de}}

\date{Received / Accepted}

\abstract{We present results and analysis for the 2\,m spectroscopic part of
the MultiSite Spectroscopic Telescope (MSST) campaign undertaken in
May/June 2002. The goal of the project was to observe the pulsating
subdwarf~B star PG\,1605+072 simultaneously in velocity and photometry
and to resolve as many of the $>$50 known modes as possible, which
will allow a detailed asteroseismological analysis. We have obtained
over 150 hours of spectroscopy, leading to an unprecedented noise
level of only 207\,m\,s$^{-1}$. We report here the detection of 20
frequencies in velocity, with two more likely just below our detection
threshold. In particular, we detect 6 linear combinations, making
PG\,1605+072 only the second star known to show such frequencies in
velocity. We investigate the phases of these combinations and their
parent modes and find relationships between them that cannot be easily
understood based on current theory. These observations, when combined
with our simultaneous photometry, should allow asteroseismology of
this most complicated of sdB pulsators.
}

\titlerunning{MSST spectroscopy of PG\,1605+072}
\authorrunning{S.~J. O'Toole et al.}

\maketitle

\section{Introduction}

Hot subdwarf B (sdB) stars are generally associated
with models of extreme Horizontal Branch (EHB) stars, i.e.\ they burn
helium in their cores, and have hydrogen envelopes that are too thin
to sustain nuclear fusion \citep[e.g.\ ][]{Heber86}. Canonical EHB
models show the stars to have a very narrow mass range around
$\sim$0.48$M_\odot$, and envelope masses $<$0.02$M_\odot$
\citep{DRO93}. How an sdB arrives at this configuration is one of the
many unresolved questions regarding these stars. One of the most
promising formation mechanisms is binary interaction or the merger of
two low-mass white dwarfs \citep[e.g.][]{HanII}. 

The discovery of short-period ($P$\,$<$\,10\,min) multimode pulsations
in some sdBs provides an excellent opportunity to probe the
interiors of these stars using the tools of asteroseismology, thereby
answering some of the above questions. Using small
telescopes (typically 1--2\,m) and examining hundreds of sdBs has
yielded around 30 of these pulsators, each with amplitude
$<$\,50\,mmag \citep[for a review of these objects, now known as V361\,Hya
  stars, see][]{Kilkenny2002}. The periods suggest that the stars are $p$-mode
pulsators, but asymptotic theory cannot be applied to the analysis of their
frequency spectra. The possibility of using oscillations to probe the
interiors of sdBs received another boost after the discovery of pulsations
with periods of 45\,min--2\,hr \citep{GFR03}. The much longer period
pulsations found in these stars indicate they are $g$-modes. The stars
are typically cooler than the $p$-mode pulsators. Theoretical
modelling has found that the pulsations in both groups may be driven by an
opacity bump due to ionisation of iron (and other iron-group elements)
\citep{CFB01,FBC03}.

There has been some limited success in asteroseismic studies of
pulsating sdBs. \citet{BFB01} detected 13 pulsation frequencies in
PG\,0014+067, matched these with a model and subsequently inferred the
star's fundamental stellar parameters. The models used
require that modes with $l=3$ need to be excited to explain
the density of modes observed. Geometric cancellation was believed to
exclude observations of such high degree modes in intensity, although
recently \citet{JDM04} claimed $l=4$ modes using high-speed, multicolour
photometry on the William Herschel telescope. Another way
to test this idea is by measuring stellar surface motions, which suffer less
from cancellation. An attempt to do this has been made for
several stars, including PG\,1605+072 (see below for more details),
PB\,8783 and KPD\,2109+4401 \citep{JP00}, KPD\,1930+2752 \citep{WJP02b},
PG\,1336-018 \citep{WJP03} and PG\,1325+101 \citep{TOe04}. Until now,
though, it has not been possible to measure enough velocity and intensity
amplitudes simultaneously to test whether $l=3$ modes in sdBs can be measured
in intensity.

PG\,1605+072 was discovered to pulsate by \citet{ECpaperVII}. The star
has the longest periods (up to 9\,min) and the highest amplitudes
(up to $\sim$60\,mmag) of the sdBV (or V361\,Hya) stars currently
known. It was clear from the discovery observations that the star
showed great asteroseismological potential, so a
multisite photometric campaign was organised to observe the star
continuously for two weeks. The results can be found in
\citet{ECpaperX}, who detected more than 50 frequencies in the
amplitude spectrum. It is difficult to say how many of these are
independent normal modes of oscillation as there is evidence for
amplitude variability of timescales of at most a few
months. \citet{Kawaler99} found a possible model match to the 5
highest amplitude frequencies measured by \citeauthor{ECpaperX},
suggesting that the star has an equatorial rotation velocity of
130\,km\,s$^{-1}$. This was given support
from spectral analysis by \citet{HRW99}, who determined $v\sin
i$=39\,km\,s$^{-1}$ from high-resolution spectroscopy, suggesting an
inclination angle of $\sim$17$^\circ$. This makes PG\,1605+072 the most
rapidly rotating and apparently single sdB known. \citeauthor{HRW99} also
found $T_\mathrm{eff}$=32\,300\,K, $\log g$=5.25, log(He/H)=$-$2.53 and
measured metal abundances; the low $\log g$ implies the star is also quite
evolved, and has moved away from the EHB.
\begin{table}[htbp]
\caption{Table of MSST spectroscopic observations. KPNO = Steward
  2.3\,m on Kitt Peak; LS = Danish 1.54\,m on La Silla; SSO = 2.3\,m
  on Siding Spring;   LP = Nordic Optical Telescope (2.56\,m) on La Palma.}
\label{tab:obs}
\begin{center}
\begin{tabular}{lcccc}
UT Date & Where & UT & UT & No. of \\
 & & start & finish & spectra \\
\hline
19/05/2002 & KPNO & 03:58 & 11:12 & 344 \\
20/05/2002 & KPNO & 04:04 & 11:40 & 434 \\
21/05/2002 & KPNO & 03:35 & 11:18 & 440 \\
22/05/2002 & LS & 02:30 & 08:43 & 315 \\
22/05/2002 & KPNO & 03:24 & 11:19 & 455 \\
23/05/2002 & LS & 02:15 & 08:35 & 348 \\
23/05/2002 & KPNO & 03:28 & 11:25 & 452 \\
24/05/2002 & KPNO & 03:16 & 11:29 & 470 \\
25/05/2002 & SSO & 14:56 & 17:56 & 160 \\
26/05/2002 & SSO & 10:43 & 17:15 & 360 \\
27/05/2002 & SSO & 11:14 & 17:56 & 360 \\
\hline
\hline
13/06/2002 & LS & 00:54 & 07:11 & 202 \\
14/06/2002 & LS & 00:36 & 07:04 & 334 \\
15/06/2002 & LS & 00:44 & 06:51 & 340 \\
16/06/2002 & LS & 00:32 & 06:45 & 318 \\
17/06/2002 & LS & 00:56 & 06:50 & 334 \\
18/06/2002 & LP & 21:36 & 04:34 & 460 \\
18/06/2002 & LS & 00:32 & 06:48 & 317 \\
19/06/2002 & LP & 20:53 & 03:14 & 400 \\
19/06/2002 & LS & 00:36 & 06:43 & 326 \\
20/06/2002 & LP & 20:54 & 04:28 & 520 \\
20/06/2002 & LS & 00:24 & 06:39 & 333 \\
21/06/2002 & LP & 20:52 & 04:26 & 500 \\
21/06/2002 & LS & 01:38 & 06:20 & 260 \\
22/06/2002 & LP & 22:12 & 04:21 & 435 \\
22/06/2002 & LS & 00:27 & 06:31 & 326 \\
23/06/2002 & LP & 21:43 & 04:18 & 435 \\
24/06/2002 & LP & 20:48 & 03:09 & 430 \\
25/06/2002 & LP & 20:49 & 04:25 & 540 \\
\hline
\textbf{total} & & & & 10892 
\end{tabular}
\end{center}
\end{table}

\begin{table*}[htbp]
\caption{Instrumentation used in the 2\,m spectroscopy part of the
  MSST campaign.}
\label{tab:inst}
\begin{center}
\begin{tabular}{lcccccccccc}
Observatory & $D$ & Grating & dispersion & slit & resolution & CCD size &
binning & $\lambda$ range & $t_{\mathrm{exp}}$ & $t_{\mathrm{dead}}$ \\
 & (m) & (lines/mm) & (\AA/pix) & (``) & (\AA) & (pix) & & (\AA) & (s) & (s) \\
\hline
Steward/KPNO & 2.3 & 832 & 0.707 & 1.5 & 1.8 & 1200$\times$800 & 1$\times$3
& 3686--4534 & 45 & 12 \\
Siding Spring & 2.3 & 600 & 1.1 & 2.0 & 2.2 & 1752$\times$532 &
1$\times$1 & 3647--5047 & 45 & 10 \\
Danish/ESO & 1.5\rlap{4} & DFOSC\#6 & 1.6\rlap{5} & 1.5 & 5.9 &
2K$\times$4K & 3$\times$2 & 3648--5147 & 35/45 & 17 \\
NOT & 2.5\rlap{6} & ALFOSC\#3 & 2.3 & 0.7\rlap{5} & 9.3 & 2K$\times$2K &
3$\times$2 & 3039--6669 & 35 & 17 \\
\hline
\end{tabular}
\end{center}
\end{table*}

Because of its long periods, high amplitudes and relative brightness,
PG\,1605+072 is an ideal target for rapid spectroscopy in comparison
with other pulsating sdBs. Velocity
variations were first detected by \citet{OBK00b}, at frequencies
corresponding to those found in photometry and with amplitudes of
up to 14\,km\,s$^{-1}$. A more detailed analysis of these and
follow-up observations was presented by \cite{OBK02}.
Other studies of PG\,1605+072 include \citet{WJP02a}, who used 4\,m
class telescopes to achieve better velocity accuracy than both
\citet{OBK00b} and \citet{OBK02}, but were hampered by timing problems
and a short temporal baseline. A study of Balmer
line indices by \citet{OJK03} measured effective temperature and
gravity changes, producing a new set of observables that may be used
for mode identification. Finally, as a feasibility study, \citet{Falter03}
conducted the first simultaneous multi-colour photometry and velocity
observations. These studies suffered from either no (or little)
simultaneous photometry, poor frequency resolution, or a combination of
both. With this in mind, we organised a multisite coordinated
spectroscopic campaign to observe PG\,1605+072 with medium resolution
spectrographs on 2\,m and 4\,m class telescopes and for as long as
possible. To complement these observations, we obtained simultaneous
photometry using 1\,m class telescopes. The project was greatly aided
by observations during the Xcov22 campaign of the Whole Earth Telescope.

In this paper we present first spectroscopic results from the
MultiSite Spectroscopic Telescope (MSST) campaign which observed
PG\,1605+072 in May/June 2002 \citep[see ][]{HDS03}. We have obtained 151
hours of time-resolved spectroscopy which was taken over 38 nights
using four 2\,m telescopes. We increase the number of detected
frequencies in velocity by a factor of two. The work represents the first step
in understanding one of the most complicated sdB pulsators. Two further papers
will be published on observations from this campaign: results from
4\,m telescopes (spectroscopy) and results from photometry (both in
preparation).

\section{Observations}
\label{sec:obs}

The MSST spectroscopic campaign was divided into two parts, shown
in Table \ref{tab:obs}. The first part (19--27/05/2002) overlapped with
the photometric observations of the MSST campaign, while the second
part (13--25/06/2002) took place about 2 weeks afterwards. We were granted time
on four 2\,m-class telescopes: the Steward Observatory 2.3\,m on Kitt Peak (6
nights); the Danish 1.54 at La Silla (8 nights in ESO time and 10 nights in
Danish time); the 2.3\,m Advanced Technology Telescope at Siding Spring
Observatory (8 nights); and the 2.56\,m Nordic Optical Telescope at La Palma
(8 nights).

Previous spectroscopic observations of PG\,1605+072 with 2\,m telescopes
proved the feasibility of a study of velocity variations, so the main focus of
this part of the MSST was to maximise the length and improve the coverage
compared with studies such as \citet{OBK02}.

\subsection{Instrumentation}
\label{sec:instrument}

In Table \ref{tab:inst} we give the details of the instrumental setup
for each telescope. To ensure accurate timing, multiple shutter tests
were conducted at KPNO and SSO. This was a necessary check since the UT
written in the image header does not correspond to the time the
shutter actually opened for some systems. At the former telescope, correction of
the UT in the image headers was required, giving an accuracy to within a
few tenths of a second, while at the latter, no corrections were
required as the times were accurate to better than one
second. From our previous experience at La Silla and La Palma we know
the timing of these telescopes to be accurate to less than one
second. To improve readout times, the CCDs were windowed. This could
not improve the size of the dead time of DFOSC and ALFOSC
however. They are large because before each exposure the systems
perform checks of both the filter and grism wheels. An override of
this feature does not seem to be possible.

\subsection{Reductions}
\label{sec:red}

The observations described above were reduced with
\textsc{iraf}, using the standard bias subtraction, flat fielding,
aperture tracing, sky correction and spectrum extraction routines. 
Wavelengths were calibrated using arc spectra that were taken roughly
once per hour. This was done to minimise spectrograph drifts from the velocity
curve and therefore reduce the $1/f$ noise component in the final amplitude
spectrum (although see below).  At the NOT arc spectra were taken
at the beginning and end of each night only, and drifts needed to be
removed by hand.

\begin{figure*}
\vspace{18cm}
\begin{center}
\includegraphics{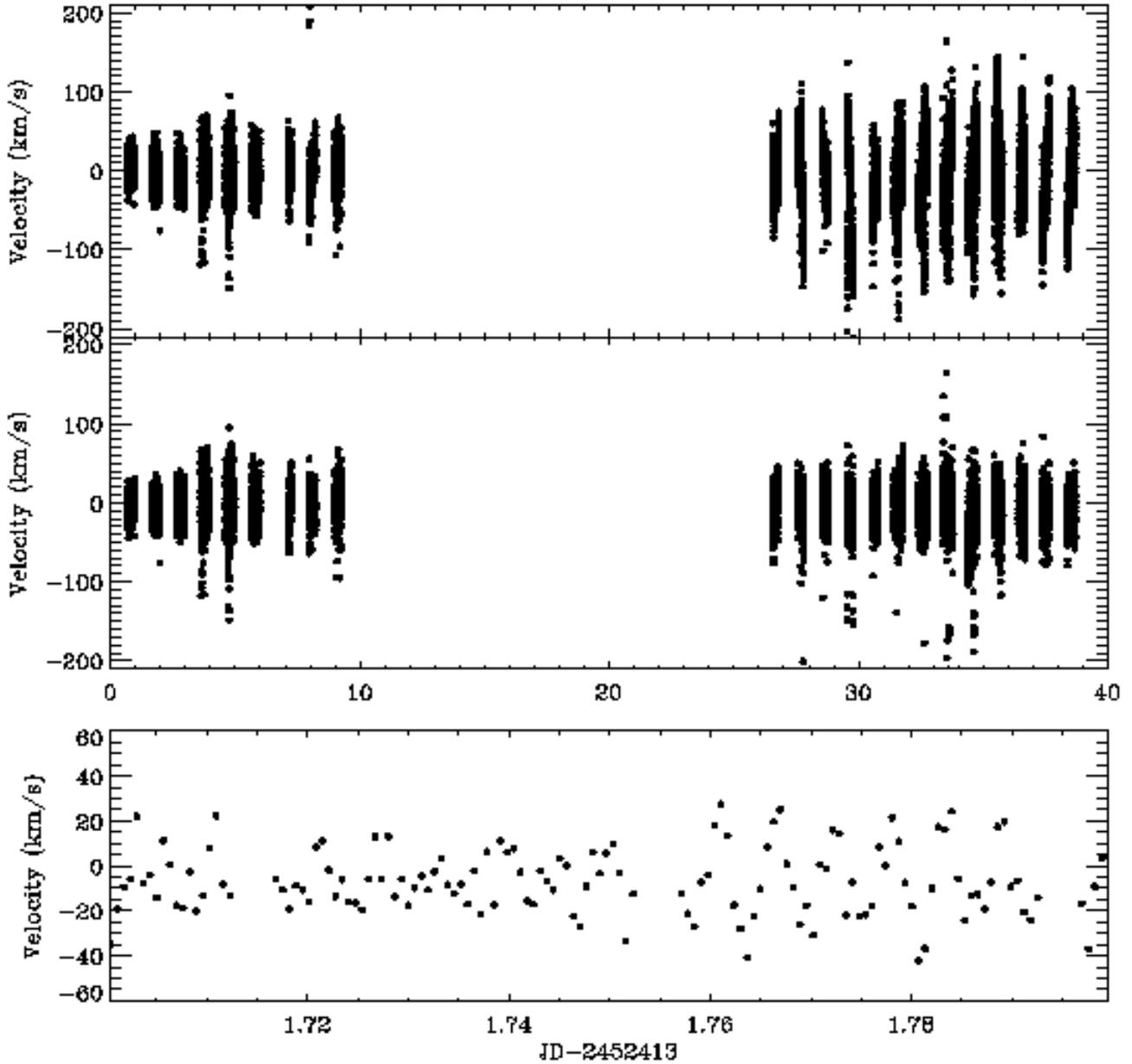}
\end{center}
\caption{Velocity curve of the entire 2\,m telescope part of the MSST
  campaign before (\emph{top}) and after (\emph{middle}) removal of slow
  trends. A section of the curve from the second night of observations
  (20 May 2002) at Steward Observatory is also shown (\emph{bottom});
  the variations at $\sim$480\,s are evident around 1.76 days.}
\label{fig:velcurve}
\end{figure*}

All velocities were determined by cross-correlating each spectrum with a
nightly template, created by median filtering that night's
spectra. For observations from the Steward Observatory 2.3\,m, the templates
were corrected to rest wavelengths using an iterative procedure,
however this was not done for the other observations since it was
deemed that the actual velocity shift is smaller than the spectral
resolution; the effect of not making this correction does not appear
to be significant. The cross correlations were done in \textsc{iraf}
using a double-precision version of the package
\texttt{fxcor}\footnote{available at
  \texttt{http://iraf.noao.edu/scripts/extern/rvx.pl}}. Unlike
previous velocity studies \citep[e.g.\ ][]{OBK02}, where Balmer lines
were examined individually, with the resulting velocity curves
combined, the entire spectrum was cross-correlated in order to achieve
the highest accuracy.

\subsection{Velocities}
\label{sec:vel}

Our resulting velocity curve is shown in the top panel of Figure
\ref{fig:velcurve}. In the case of DFOSC and ALFOSC, both
grism spectrographs, regular arc spectra reduce the drift, but by no means
remove it completely, which is seen as apparently high scatter, particularly
on the right side of the figure. In contrast, the drift in the
velocities determined from data taken with grating spectrographs (Steward and
SSO 2.3\,m telescopes) is almost completely removed. Similar drifts are seen
in other time-series ALFOSC spectra \citep[e.g.\ observations of another V361
  Hya star, PG\,1325+101 by ][]{TOe04}. The drifts, and their removal, are
discussed in more detail below.

\begin{figure}
\vspace{6cm}
\begin{center}
\includegraphics{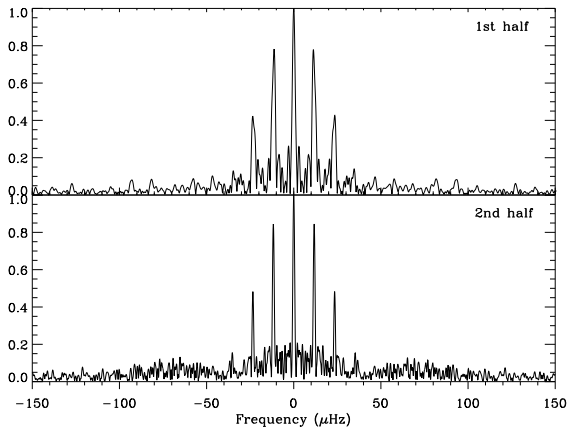}
\end{center}
\caption{Spectral windows of the two halves of the 2\,m spectroscopic
  campaign. The top panel represents the window for 19-27/05/2002,
  while the bottom represents the window for 13-25/06/2002.}
\label{fig:sw2halves}
\end{figure}

\begin{figure}
\vspace{6cm}
\begin{center}
\includegraphics{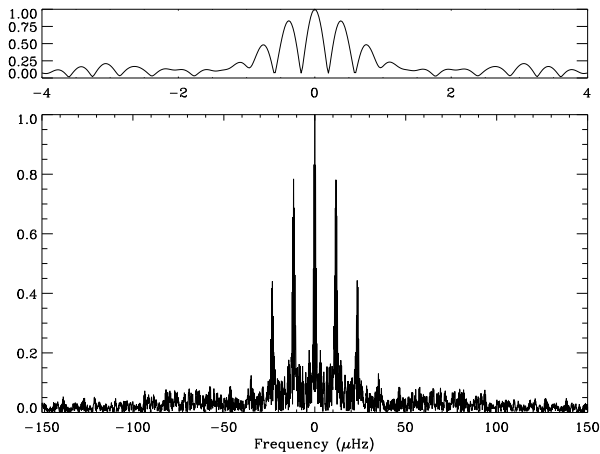}
\end{center}
\caption{Spectral window of all 2\,m spectroscopic observations. The
  top panel shows fine structure in the central peak caused by the 2
  week gap in observations. The spacing between the alias peaks is
  $\sim$0.38\,$\mu$Hz.}
\label{fig:swcomb}
\end{figure}

The spectral windows of the two parts of the campaign and the whole campaign
are shown in Figures \ref{fig:sw2halves} and \ref{fig:swcomb},
respectively. These serve as a guide to the aliasing we can expect in the
amplitude spectrum. It can be seen from Figure \ref{fig:sw2halves} that
the coverage during two halves of the campaign was approximately the same,
although the frequency resolution of the second part is better than the
first. The result of this is that the spectral window of the entire campaign
is more or less the same as the two halves, except for the introduction of
fine structure caused by their two week separation, which is shown in the
top panel of Figure \ref{fig:swcomb}. The separation between the main
alias peaks corresponds to one cycle-per-day (measured to be
11.67\,$\mu$Hz rather than 11.57\,$\mu$Hz), while the small separation
(shown in the inset) is $\sim$0.38\,$\mu$Hz. The latter alias
corresponds to the difference between midpoints in the two parts of
the campaign, rather than the length of time between them. These
aliases must be considered when conducting any frequency analysis.

Because the quality of the data varies from night to night, and even during
the night, we have assigned weights to each velocity based on the velocity
error determined by \texttt{fxcor}. Studies by \citet{OBK00b} and
\citet{OBK02} used the inverse square of the local rms scatter of each night
as weights -- the advantage of using the velocity error is that individual
points are considered, not just an average over the whole night. Also,
since variations are visible in many parts of the velocity curve (see
the bottom panel of Figure \ref{fig:velcurve}), the real variation may
be confused with noise. Finally, data
that lie outside $\pm$100\,km\,s$^{-1}$ -- where a sharp drop in S/N
was caused by passing clouds -- were weighted significantly lower than
those within.

Despite the regular taking of arc spectra, there are still drifts and jumps in
the velocity curve, especially in the observations made with grism
spectrographs. To remove these, we firstly did a crude analysis, whereby we
fit sinusoids to and subtracted them from each night's data until all
that remained were the white and $1/f$ noise components. To the
signal-free residuals, we fit a series of polynomials of degree 1-3,
and then subtracted the fits from the original raw velocity
curve. The trends we remove are non-periodic, and caused by
instrumental effects; they are most likely
not intrinsic to the star. The final velocity curve with trends
removed is shown in the middle panel of Figure
\ref{fig:velcurve}. This is a significant improvement on our initial
velocity curve, increasing the possibility of detecting low amplitude
oscillations. The bottom panel of Figure \ref{fig:velcurve} shows a
section of the velocity curve from the second night of observations
(20 May 2002) at Steward Observatory; velocity variations with a
period around 480 seconds are evident between 1.76 and 1.78 days.

\begin{figure*}
  \vspace{10cm}
  \begin{center}
    \includegraphics{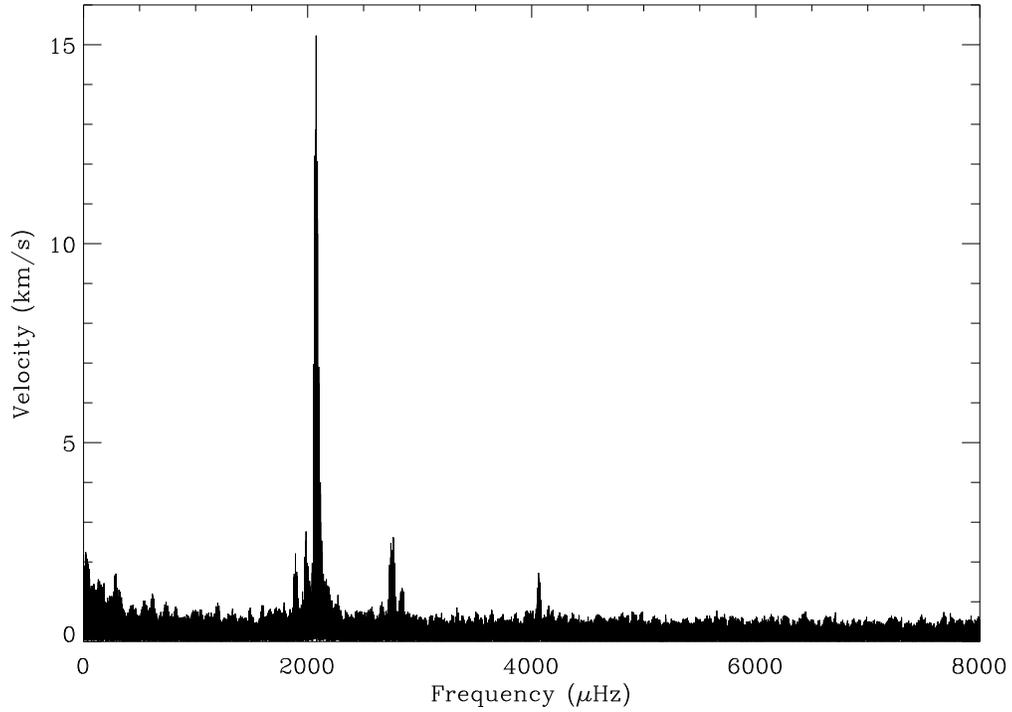}
  \end{center}
  \caption{The velocity amplitude spectrum based on 2\,m telescope
  spectroscopy of PG\,1605+072. The $1/f$ component of the noise is
  low because of our removal of slow drifts. The final white noise
  level is 207\,m\,s$^{-1}$.}
  \label{fig:aspec}
\end{figure*}

\begin{figure*}
  \vspace{10cm}
  \begin{center}
    \includegraphics{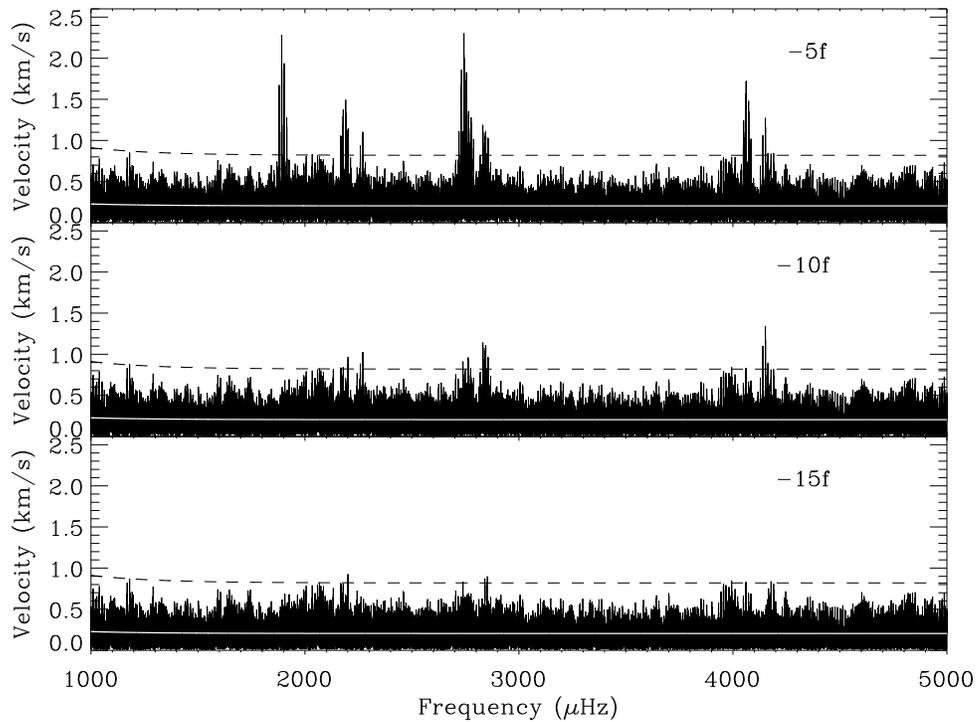}
  \end{center}
  \caption{Amplitude spectrum after prewhitening. The white line shows the
  noise function determined in section \ref{sec:analysis}, while the dashed
  line represents our detection threshold, or four times the noise function.}
  \label{fig:prewhiten}
\end{figure*}

The velocity amplitude spectrum was calculated using a weighted Fourier
transform and is shown in Fig \ref{fig:aspec}. We determined the noise level
in the spectrum in two ways: by averaging the amplitude spectrum in regions
where no signal exists (i.e.\ at high frequencies); and by fitting an
exponential function to the prewhitened spectrum. The former method is
only an indication of the white noise level, which we find to be
207\,m\,s$^{-1}$, while the latter method includes an approximation of the
$1/f$ component as well. The noise function we derived is shown in Figure
\ref{fig:prewhiten}. Because of the removal of slow drifts, the $1/f$ noise is
virtually the same as the white noise in the frequency range of interest. The
noise level is almost 45\% lower than in \citet{OBK02}. 

\section{Frequency analysis}
\label{sec:analysis}

The frequencies presented here were determined using \texttt{Period98}, a
least-squares fitting program \citep{p98} that fits multiple sinusoids
simultaneously with the following form:
$$Z+\Sigma_{i=0}^N A_i\sin(2\pi f_it+\phi_i)$$
where $Z$ is the zero point shift, $A_i$ is the amplitude, $f_i$ is the
frequency and $\phi_i$ is the phase. 
The analysis was performed
in an iterative manner, and the amplitude of each peak was compared
with the noise function discussed earlier. If the peak
is four times the noise then we consider it to be real. There are two
exceptions to this, however: peaks at 3967.29\,$\mu$Hz and 3993.88\,$\mu$Hz
are considered real because of their very good correspondence to linear
combinations of higher amplitude modes. Our full list of frequencies is shown
in Table \ref{tab:combined}.

\begin{table}[htbp]
  \caption{Frequencies, velocity semi-amplitudes and periods from the entire
  2\,m campaign; $n$ is the amplitude rank. Entries marked with an
  asterisk (*) have amplitudes below our threshold of four times the
  white noise level, however their correspondence with combination
  frequencies is good enough to suggest their reality.}
  \begin{center}
    \leavevmode
    \begin{tabular}{cccccc}
      \hline
      $n$ & $f$ & $V$ & $P$ & Freq & $\Delta$ \\
       & ($\mu$Hz) & (km\,s$^{-1}$) & (s) & sums & ($\mu$Hz) \\
      \hline
      6 & 1891.41 & 2.322  & 528.71 & & \\
      5 & 1985.89 & 2.474  & 503.55 & & \\
      1 & 2075.80 & \llap{1}5.429 & 481.74 & & \\
      3 & 2101.91 & 2.971  & 475.76 & & \\
      2 & 2102.55 & 5.372  & 475.61 & & \\
      14 & 2201.20 & 0.977 & 454.30 & & \\
      17 & 2202.18 & 0.919 & 454.10 & & \\
      13 & 2270.37 & 0.998 & 440.46 & & \\
      18 & 2738.42 & 0.901 & 365.17 & & \\
      4 & 2743.01 & 2.497  & 364.56 & & \\
      11 & 2762.67 & 1.155 & 361.97 & & \\
      7 & 2763.50 & 2.121  & 361.86 & & \\
      9  & 2764.45 & 1.655 & 361.74 & & \\
      15 & 2842.94 & 0.967 & 351.75 & &  \\
      12 & 2844.59 & 1.112 & 351.54 & & \\
      16 & 2852.31 & 0.931 & 350.59 & & \\
      22 & 3967.29\rlap{$^*$} & 0.788 & 252.06 & $f_1+f_6$ & +0.08 \\
      21 & 3993.88\rlap{$^*$} & 0.810 & 250.38 & $f_2+f_6$ & $-$0.08 \\
      8 & 4061.70 & 1.777  & 246.20 & $f_1+f_5$ & $-$0.01 \\
      10 & 4151.62 & 1.364 & 240.87 & $2f_1$ & +0.02 \\
      19 & 4178.36 & 0.858 & 239.33 & $f_1+f_2$ & +0.01 \\
      20 & 6137.56 & 0.845 & 162.93 & $2f_1+f_5$ & +0.07 \\
      \hline
    \end{tabular}
    \label{tab:combined}
  \end{center}

\end{table}

\subsection{Below 2500\,$\mu$Hz}
\label{sec:below2500}

Most of the oscillation power of PG\,1605+072 lies in the frequency range
1800--2500\,$\mu$Hz. The dominant peak in the amplitude spectrum, as seen in
the Figure \ref{fig:aspec}, lies at 2075.80\,$\mu$Hz ($f_1$). 
In earlier velocity studies \citep{OBK00b,OBK02,WJP02a} this frequency
has been found to have a much reduced amplitude compared to that seen in
intensity by both \citet{ECpaperVII} and \citet{ECpaperX}, although the most
recent study prior to our observations found that it was dominant again
\citep{Falter03}. This apparently variable amplitude may be intrinsic to the
star, or may simply be caused by beating between two or more very closely
spaced peaks. Since these possibilities are difficult to explain
theoretically, we leave both open for further investigation.

The peaks found at 2101.90\,$\mu$Hz and 2102.55\,$\mu$Hz match quite well
those found by \citet{OBK02}, although there is a small frequency difference.
\citeauthor{OBK02} found an extra frequency at 2102.83\,$\mu$Hz, probably
because of their better frequency resolution. These frequencies were not
resolved by the other three velocity studies of PG\,1605+072. The multisite
photometric study of \citet{ECpaperX} found the two peaks, but at slightly
different frequencies, and different again to the frequencies measured by
\citet{OBK02}. The reason for these apparent frequency differences, as well as
the variation of the 2075.8\,$\mu$Hz peak, is
uncertain, although they resemble the stochastically excited
oscillations found in the Sun and solar-like stars. A recent study by
\citet{PL05} seems to rule this out however. This is perhaps not surprising,
since solar-like oscillations are related to stellar convection zones in the
atmosphere, and sdBs are not predicted to show such zones. The nature
of this amplitude variability remains unknown; it seems
likely to us that it is the result of the damping and re-excitation.

We now comment briefly on the other modes detected in this region.

\renewcommand{\labelitemi}{$\bullet$}
\begin{itemize}
\item The peak at 2202.18\,$\mu$Hz was actually detected at
  $\sim$2190.5\,$\mu$Hz, however we believed this to be separated from
  the true frequency by an alias. \citet{ECpaperX} found a peak in
  their photometry at $\sim$2201.93, so we have fit
  2190.5+11.67\,$\mu$Hz instead. 

\item \citet{ECpaperX} measured the peak at 1985\,$\mu$Hz to have the
  8th highest amplitude, however, in subsequent studies its amplitude
  relative to the other peaks is consistently higher. For example,
  \citet{WJP02a}, \citet{OBK02} and \citet{Falter03} detected the
  mode, while \citet{OBK00b} did not.

\item The 1891\,$\mu$Hz peak was detected by \citet{OBK00b},
  \citet{WJP02a} and \citet{OBK02}, but not by \citet{Falter03},
  although in Figure 3 of the latter study, the peak is possibly
  present, but the low frequency resolution and S/N makes it difficult
  to be certain.

\item The mode at $\sim$2270\,$\mu$Hz has only previously been
  marginally detected by \citet{OBK02} -- our detection is much more
  confident. This mode may form part of a rotationally split triplet
  \citep[][ see Section \ref{sec:future}]{Kawaler99}.
\end{itemize}

There is still an excess of power in the 2000-2200\,$\mu$Hz range, in
particular around $\sim$2085\,$\mu$Hz and $\sim$2167\,$\mu$Hz. An
excess also exists around 1650\,$\mu$Hz, however no pulsation frequency has
ever been measured in that range for PG\,1605+072 before. While we are
certainly not claiming these as detections, it is encouraging that the first
two frequencies correspond well to frequencies measured by
\citet{ECpaperX}. Full analysis of our photometry may clarify the
reality of these peaks, although perhaps not if they have $\ell>2$.

\subsection{2500-3500\,$\mu$Hz}
\label{sec:midrange}

The strongest mode in this range is at 2743.01\,$\mu$Hz, and was the
strongest mode overall in the studies of \citet{OBK02} and \citet{WJP02a},
although in both these studies it was split in two (by quite different
amounts). This has already been discussed by \citeauthor{OBK02}\, No such
splitting is seen in our data, suggesting again that the mode amplitudes are
varying with time. This idea is supported by several low amplitude modes found
nearby by \citet{ECpaperX}.

The modes around 2763\,$\mu$Hz have been previously detected in photometry,
although \citet{ECpaperX} found them only after observations additional to
their main campaign were included in the analysis. Otherwise they list two
frequencies at 2761.3 and 2766.3 in their Table 3. A peak at 2764.1\,$\mu$Hz
was detected by \citet{ECpaperVII} which matches our detection well
considering their very complicated window function. As before, these modes
could be the result of damping and re-excitation of one pulsation mode.

We have detected two peaks around 2843\,$\mu$Hz for the first time in
velocity, although they have been observed before by
\citet{ECpaperX}. The lower frequency of the pair was originally
detected at 2831.28\,$\mu$Hz, however because of the
\citeauthor{ECpaperX} detection separated by approximately one
cycle-per-day, we believe this to be the correct value. These
frequencies may form part of a multiplet (see Section \ref{sec:future}). 

There are two frequencies we have detected at 2738.42\,$\mu$Hz and
2852.31\,$\mu$Hz, which correspond approximately to peaks found by
\citet{ECpaperVII} in photometry, but not seen by \citet{ECpaperX}. Again, a
full analysis of our simultaneous photometry should help to clarify
the status of these peaks.

\subsection{Above 3500\,$\mu$Hz}
\label{sec:above3500}

In this range the frequencies we have detected are linear
combinations of the frequencies below 2500\,$\mu$Hz discussed in Section
\ref{sec:below2500}. 

We have detected the first harmonic of the highest amplitude peak in
our observations ($f_1$) ; this is the first detection of this kind in
velocity, although \citet{TCvK03} measured a combination frequency in
a velocity study of the DAV white dwarf G29-38. \citet{ST99}
considered line-profile variations due to non-radial adiabatic
pulsations in rotating stars. They found that any oscillations where
temperature effects dominate are not expected to show detectable harmonic
variability in the line profile at all. Our detection therefore suggests that
line profile variations play a significant role. More theoretical work
is needed in this area.

The five other frequencies above 3500\,$\mu$Hz found in Table
\ref{tab:combined} are linear combinations of the strongest modes. 
This is first time that the second order linear combination $2f_1+f_5$
has been detected in PG\,1605+072, although its frequency
(6137.56\,$\mu$Hz) is not the highest that has been detected in this
star. \citet{ECpaperVII} detected frequencies up to
$\sim$8400\,$\mu$Hz, although these frequencies were not detected in
the multisite observations of \citet{ECpaperX}. All other combination
frequencies are first order, and have been detected in photometry by
earlier studies.

\begin{table}[htbp]
\caption{Relationships between phases of combination frequencies and their
  parent modes. Phases have been normalised to the range [$-\pi,\pi$].}
\label{tab:combphases}
\begin{center}
\begin{tabular}{ccccc}
$f_i+f_j$ ($f_k$) & $\phi_i$ & $\phi_j$ & $\phi_k$ & Relation \\
 & ($^\circ$) & ($^\circ$) & ($^\circ$) & \\
\hline
$2f_1$ ($f_{10}$) & \llap{$-$}35$\pm$1 & --- & 94$\pm$9 &
$2\phi_1\approx\pi+\phi_{10}$ \\
$f_1+f_2$ ($f_{19}$) & \llap{$-$}35$\pm$1 & 31$\pm$2 & 164$\pm$13 &
$\phi_1+\phi_2$ \\
 & & & & $\approx\pi+\phi_{19}$ \\
$f_1+f_5$ ($f_8$) & \llap{$-$}35$\pm$1 & \llap{1}58$\pm$5 & 57$\pm$7 &
$\phi_1+\phi_5$ \\
 & & & & $\approx\pi-\phi_{8}$ \\
$2f_1+f_5$ ($f_{20}$) & \llap{$-$}35$\pm$1 & \llap{1}58$\pm$5 & $-$169$\pm$14 &
$\phi_1-\phi_5\approx-\phi_{20}$ \\
$f_1+f_6$ ($f_{22}$) & \llap{$-$}35$\pm$1 & \llap{1}56$\pm$5 & $-$121$\pm$15 &
$\phi_1+\phi_6\approx-\phi_{22}$ \\
$f_2+f_6$ ($f_{21}$) & 31$\pm$2 & \llap{1}56$\pm$5 & $-$170$\pm$14 &
$\phi_2+\phi_6\approx\phi_{21}$ \\
\hline
\end{tabular}
\end{center}
\end{table}

When we examine the phases of the combination frequencies and their
parent modes, we find a few interesting results, although we remain cautious
about any possible interpretation. The relationships are shown in Table
\ref{tab:combphases}. Phase errors are determined in complex phase
space. What these relations say about the character of
the modes is unclear without theoretical modelling. \citet{Wu2001}
investigated combination frequencies in pulsating DA and DB white
dwarfs and derived analytic expressions for their amplitudes and
phases. Wu also found that mode identification is possible using
combination frequencies. Unfortunately we cannot apply the results of
her analysis here, since pulsating white dwarfs have convection zones
while sdBs do not, but the potential is exciting nevertheless.

\section{Further discussion}
\label{sec:future}

Because of the large number of modes observed in PG\,1605+072 and its
relative brightness, it has become one of the most studied sdB
stars.

\citet{Kawaler99} found a model
matching the 1891, 2075 and 2270\,$\mu$Hz peaks found by
\citet{ECpaperX} to an $l=1$ triplet. From this he derived a rotation
velocity of $\sim$130\,km\,s$^{-1}$; \citet{HRW99} later measured
$v\sin i=39$\,km\,s$^{-1}$ from high resolution spectroscopy, a value
that is consistent with Kawaler's determination if the inclination
angle of the star is $\sim17^\circ$. We have observed each of these
modes in our velocity amplitude spectrum for the first time. 

Interestingly, in Table \ref{tab:combined} there are three
pairs of frequencies -- (2101.90, 2201.20)\,$\mu$Hz, (2102.55,
2202.18)\,$\mu$Hz and (2742.99, 2842.95)\,$\mu$Hz -- separated by
$\sim$100\,$\mu$Hz. There are further pairs that can
be seen in \citet{ECpaperX} at (1985.32, 2085.84)\,$\mu$Hz and
(2842.05, 2942.77)\,$\mu$Hz. The first peak of the former pair is present in
our amplitude spectrum and as discussed in Section \ref{sec:below2500}, there
is evidence for a low-amplitude peak at $\sim$2085\,$\mu$Hz. \emph{One must be
careful not read too much into these patterns without analysis of our
photometry, however!}

\section{Summary and Conclusions}
\label{sec:sum}

We have presented observations and analysis of the 2\,m spectroscopic
observations of the pulsating sdB star PG\,1605+072 from the MSST
campaign. With more than 150 hours and almost 11\,000 spectra, we have
been able to reduce our white noise level to 207\,m\,s$^{-1}$,
allowing the detection of the largest number of frequencies in
velocity ever in an sdB. Among these are five first order and one
second order linear combinations; this is only the second time that
such frequencies have been observed in velocity. The highest amplitude
mode from the discovery observations is once again dominant, after
several seasons of low amplitudes or even non-detections.

Since such a large amount of information can be extracted from this
data set -- for example a line index study similar to \citet{OJK03} --
we have decided to make it available for public use (please contact
the authors). This will
also eventually include our 4\,m spectroscopy and our almost 400 hours
of photometry. With the combination of all of these data, asteroseismology
may begin to unravel one of the most complex pulsating stars known.

\begin{acknowledgements}
We are grateful to the various time allocation commitees who gave generously. 
SJOT was partially supported by an Australian Postgraduate Award and is
currently supported by the Deutsches Zentrum f\"ur Luft- und Raumfahrt (DLR)
through grant no.\ 50-OR-0202. VMW was supported by the UK Particle Physics
and Astronomy Research Council (PPA/G/S/1998/00019 and
PPA/G/O/2001/00068). CSJ acknowledges support to the Armagh Observatory from
the Northern Ireland Dept. of Culture, Arts and Leisure.
We would also like to thank Tim Bedding for kindly providing support and
funding for observers to stay at SSO.

\end{acknowledgements}

\bibliographystyle{aa}

\end{document}